\documentstyle[12pt]{article}

\def\be{\begin{equation}}
\def\ee{\end{equation}}
\def\bea{\begin{eqnarray}}
\def\eea{\end{eqnarray}}
\def\nn{\nonumber}
\def\nsp{\noindent}

\renewcommand{\thefootnote}{\fnsymbol{footnote}}

\begin{document}
\begin{titlepage}
\begin{center}
\hfill {\it DOE/ER/40762-061}\\
\vspace{-.1in}
\hfill {\it U. of MD PP\#96-001}\\
\vspace{-.1in}
\hfill {\it TAN-FNT-95-07}\\
\vspace{-.1in}
\hfill hep-ph/9507328\\

\vspace*{2.0cm}
{\large\bf NUCLEON ELECTRIC POLARIZABILITY IN SOLITON
MODELS AND THE ROLE OF THE SEAGULL TERMS}

\vskip 1.5cm

Norberto N. SCOCCOLA$^{a,b}$
\footnote[2]{Fellow of the CONICET, Buenos Aires, Argentina.}
and Thomas D. COHEN$^c$
\footnote[3]{On leave from Department of Physics, University
of~Maryland,
College~Park, MD~20742.}

\vskip .2cm
{\it
$^a$ INFN, Sez. Milano, Via Celoria 16, 20133 Milano, Italy.\\
$^b$ Departamento de F{\'{\i}}sica,
Comisi\'on Nacional de Energ\'{\i}a At\'omica,\\
\vspace{-.12in}
Av. Libertador  8250, 1429 Buenos Aires, Argentina.\\
$^c$ Department of Physics and Institute for~Nuclear Theory,\\
\vspace{-.12in}
University of Washington, Seattle, WA~98195, USA.}
\vskip 1.cm
July 1995
\vskip 1.25cm
{\bf ABSTRACT}\\
\begin{quotation}
The full Hamiltonian of the soliton models contains
no electric seagull terms.  Here it is shown that if one
restricts the fields
to the collective subspace then   electric seagull terms are
induced in the effective Hamiltonian. These effective seagull
contributions are consistent with gauge
invariance.  They also reproduce the leading nonanalytic behavior
of a large $N_c$ chiral perturbation theory calculation of the
electric polarizability.
\end{quotation}
\end{center}
\end{titlepage}

\renewcommand{\thefootnote}{\arabic{footnote}}

\newpage

\section{Introduction}

The electric and magnetic polarizabilities  of the nucleon are
important low energy observables.  During the past several years
there has been significant improvements in the experimental
measurement of these quantities\cite {Fed91}.  The ability to
explain the polarizability is a significant test of any proposed
model of the nucleon.  Of course, in order to use the
experimental
data to discriminate between the various models on the market,
one needs to be able to calculate the polarizabilities within
each
of the models.  This is not necessarily
trivial either technically or  conceptually.
This paper concerns a central conceptual issue in calculations of
the electric polarizability in the topological soliton models
\cite{Sky61,ZB86}
and in other hedgehog soliton or bag models based on large $N_c$
QCD such
as the chiral-quark meson soliton model \cite{BB84},
the chiral or hybrid bag model \cite{VJ90}
or the soliton approach to the Nambu--Jona-Lasinio (NJL)  model
\cite{NJL61}. As all these models behave identically with
regard
to the main issues in this article, for simplicity we will refer
to all of them as soliton models.

There have been a number of calculations of the nucleon electric
polarizability in soliton models \cite{Che87}--\cite{GS93}.
In all of these calculations the dominant contribution to the
polarizability has been a seagull contribution---{\it i.e.}, a
term in
the effective Hamiltonian proportional to the square of the
external electric field.  In fact, the derivation of this
seagull term in all of the calculations
to date has been rather naive.  The method used was to couple the
Lagrangian density for the model to an external electromagnetic
potential in the usual way.  An external electric
field in the $\hat{z}$ direction is introduced by choosing
$A_0 = - E z$.  A collective Lagrangian is obtained by inserting
a rotating hedgehog solution into the Lagrangian density and
integrating over the spatial coordinates; the collective
variables being the angles specifying
the rotation and  their associated angular velocities.  Finally,
the
collective Hamiltonian is obtained from the collective
Lagrangian.
The seagull term in the collective Hamiltonian is then simply
identified as minus the seagull contribution
to the collective Lagrangian since this term contains no time
derivatives.

\clearpage

This procedure is essentially the one used to obtain a
collective Hamiltonian in the absence of external fields.
However, in the present context such a procedure seems quite
cavalier in view of the fact that when the Lagrangian contains
derivative couplings the identification of the
part of the Lagrangian lacking time derivatives with minus
the corresponding terms in the Hamiltonian is,
in general, not correct.  For example, in scalar
electrodynamics there is  an electric seagull term in the
Lagrangian
given by
${\cal L} =  \phi^* \phi A_0^2 $,
while in $\cal{H}$ there are no terms proportional to
$A_0^2$ \cite{IZ}.  Indeed, it is obvious that the existence of
such a term
in the Hamiltonian would violate gauge invariance.

It has recently been argued that the procedure used in obtaining
the electric seagull terms in the collective Hamiltonian of the
Skyrme model and other soliton models is more than
cavalier; namely, that
it is wrong.  A general form of this  argument is due to
L'vov \cite{Lvo93} who argues that a general theorem of Brown
\cite{Brown} shows that existence of electric seagulls violates
local gauge invariance.  A particularly
transparent  version of this argument is that if the procedure
used to obtain electric seagulls in the Skyrme model
were used with a constant $A_0$ then one
would obtain an energy of the soliton directly proportional to
$A_0^2$. This would be manifestly absurd since the energy of a
charged
state in an external constant potential has an electric energy
shift directly proportional to $A_0$ with no higher
powers \cite{Lvo_pc}.  Moreover,  Saito and Uehara \cite{SU94}
explicitly demonstrated that the full Hamiltonian of the Skyrme
model coupled to an external electric
potential has no  seagulls.

The arguments in Refs. \cite{Lvo93,SU94}
appear to be compelling. Therefore it would seem
that the electric seagulls should be absent and the calculations
in Refs. \cite{Che87}--\cite{GS93}
are simply incorrect.  On the other hand, there is  quite strong
evidence that the electric seagull terms as calculated in these
works are correct.  This evidence comes
from the behavior of the electric polarizability $\alpha$ as one
approaches the large  $N_c $ and chiral limits (with the
$N_c \rightarrow \infty$ limit taken first).  The polarizability
in this limit is completely determined via large $N_c$ chiral
perturbation and is given by\footnote{This result is precisely
three times the naive chiral perturbation result of
Ref. \cite{BKM91}.  This factor of three
comes from the contribution of $\Delta + \pi$ intermediate states
which contribute in the large $N_c$ limit since the $\Delta$
becomes degenerate with the nucleon and gives rise to new
infrared
behavior.  In this limit the inclusion of the $\Delta$
contributions always yields three times the naive result of
chiral perturbation theory \cite{BC93,CB92}.}
\begin{equation}
\lim_{ \begin{array}{c}
{\scriptstyle m_\pi \rightarrow 0} \\[-1.mm]
{\scriptstyle N_c \rightarrow \infty }
\end{array}}  \,
\alpha
= \frac{5 \, e^2
\, g_A^2}{128 \, \pi^2 \, f_\pi^2 \, m_\pi} \label{chipt1}
\end{equation}

Now consider the electric seagull contributions to $\alpha$
using the methods of Refs. \cite{Che87}--\cite{GS93}.
In general, the
result is given by an integral which depends on the details of
the model.  However, as $m_\pi$ goes to zero
the integral is  increasingly dominated by the long range tail.
The overall strength of the contribution is proportional to the
square of the amplitude of the pion field which in the standard
treatment of soliton models is directly proportional to
$g_A^2/f_\pi^2$ (where $g_A$ is the appropriate value of $g_A$
for the model in question). Evaluating the integral for $\alpha$
in this limit precisely reproduces Eq.~(\ref{chipt1}). Moreover,
apart from the seagull, no contribution to $\alpha$ in the
calculations based on soliton models diverges as $m_\pi
\rightarrow 0$.
Indeed it is precisely for this reason that seagulls dominate the
result when $m_\pi$ is  finite but small as it is in the physical
world.
Thus, the electric seagull terms appear to be required to get the
correct behavior in the large $N_c$ and chiral limits.

Thus there is a paradox: On the one hand,
there is both a general argument based on gauge invariance and an
explicit calculation showing the absence of seagulls in the
Hamiltonian; while on the other hand, these seagulls are
apparently
necessary to get the correct result in the large $N_c$ and chiral
limits.
The purpose of the present paper is to
resolve this paradox.

The resolution is, in fact, quite simple.
The full Hamiltonian for the soliton models does not contain any
electric seagulls as is argued in Refs. \cite{Lvo93,SU94}.
However, the model is studied in the large $N_c$ limit in which
there is a collective manifold of configurations
which dominate the low energy behavior.
For the original Skyrme model, this manifold is simply  the space
of rotating hedgehogs \cite{ANW83}.  For models including
explicit fermion or vector meson degrees
of freedom the collective manifold includes ``cranking'' effects
as discussed in Ref. \cite{CB86}.  The central point in resolving
the paradox is
that while the full Hamiltonian has no electric seagulls, the
{\it effective
Hamiltonian} in terms of the  collective variables can, and in
general will, have such
terms.  This will be shown explicitly below.

Moreover, it is clear why the effective
theory can have electric seagull terms without violating Brown's
theorem on electric seagulls.  That theorem depends explicitly on
the fact that the field theory must be {\it local}.  However, the
constraint to some collective manifold
will depend on spatial integrals and the
effective Hamiltonian which emerges does
not correspond to a local field theory.

However, there is still a potential problem: namely, even if
electric seagulls do exist,  the identification of the electric
seagull term in $\cal{H}^{\rm coll}$ as minus the seagull in
$\cal{L}^{\rm coll}$ is problematic.  In general, this
identification is
clearly incorrect; the case of a constant $A_0$ provides an
explicit
example where this procedure fails.  We will show, however, that
in certain cases the identification is in fact valid.  In
particular, it is valid when the term in the collective
Hamiltonian (or Lagrangian) linear in the external field
vanishes for all values of the collective variables.  One
can only expect this to happen due to symmetry.  In fact, this
will occur if all the configurations in the collective manifold
have  the same parity and the
external $A_0$ is of odd parity; which is precisely the case of
the soliton model collective manifold of states  in a constant
electric field.

This paper is organized as follows: in Sec. 2 we discuss in the
simple model of a charged scalar field the effect of introducing
a constraint in the allowed field configuration. We show that
this
automatically leads to the presence of a seagull term in the
effective
Hamiltonian. In Sec. 3 we study in all detail a soliton model
in an external $A_0$ field, namely we discuss how the electric
seagulls do contribute to the nucleon electric polarizability in
the Skyrme model. In Sec. 4 we give our conclusions.

\section{Electric seagulls for a constrained system of charged
scalars}

Before discussing the specifics of the soliton models
constrained to a collective manifold, it is useful to
study a simpler system which raises the same issues.
We will consider a system of charged scalars interacting
with an external electromagnetic field; the fields are subject to
a constraint
that they must be in some collective manifold.  Here we will
assume that the external magnetic field is zero and we will work
in a gauge in which $A_i=0$. We will denote the charge of the
particle
by $Q$.

The Lagrangian density for the system in the absence of any
constraint is
\begin{equation}
{\cal L} \, = \, \partial_{\mu} \phi^{*} \partial^{\mu} \phi \, -
\, m^2 \phi^* \phi \, - \,  i Q \, A_0 \,
(\phi^* \dot\phi \,- \,\phi \dot\phi^*)
\, + \, Q^2 \ A_0^2 \ \phi^* \phi
\end{equation}
which leads to the following Hamiltonian density
\begin{equation}
{\cal H} \, = \, \pi^* \pi \, +
\,\vec \nabla \phi^* \cdot \vec \nabla \phi
\, + \, m^2 \phi^* \phi \, + \, i Q A_0 (\pi^* \phi^* - \pi
\phi)\;.
\end{equation}
As expected, for the full theory there is a seagull in the
Lagrangian but not the Hamiltonian.

Suppose we now expand the fields in terms of some (for now
arbitrary) orthonormal and complete set of functions, $f_j(x)$,
satisfying
\begin{eqnarray}
\int \, dV \,  f_i^*(x) f_j(x) = \delta_{i, j} \;,\\ [.18in]
\sum_{i} f_i^*(x) f_i(y) = \delta^3(x-y) \;.
\end{eqnarray}
The Hamiltonian becomes
\begin{equation}
H \, = \sum_{i,j} \left[ \,\pi^*_i \delta_{i,j} \pi_j \,
+ \phi^*_i B_{i,j} \phi_j
\, + \, i Q (\pi^*_i A^*_{i,j} \phi^*_i - \pi_i A_{i,j} \phi_j)
\right]
\label{H2}\end{equation}
where $\phi_i = \int \, dV f^*_j(x) \phi(x)$ and $\pi_i = \int
\, dV f_j(x) \pi(x)$ are the dynamical variables and their
conjugate
momenta, respectively. The matrices $A$ and $B$ are
defined by the following:
\begin{eqnarray}
A_{i,j} \, & \equiv & \, \int \, dV \, A_0(x) f^*_i(x) f_j(x) \;,
\label{A}\\ [.18in]
B_{i,j} \, & \equiv & \,m^2 \delta_{i,j}
+ \int \, dV \, \vec \nabla f^*_i (x) \cdot \vec \nabla f_j(x)
\;.
\end{eqnarray}
At this stage the Hamiltonian in Eq.~(\ref{H2}) is completely
equivalent to the original Hamiltonian and, of course, there are
no electric seagulls.
Now, suppose that the dynamics were constrained so that fields
had to be in a collective space of dimension $N$.  A simple way
to
do this is to choose a particular basis set $f_j(x)$ in which to
expand and then to impose the following condition on the field
variables:
\begin{equation}
\phi_k = 0 \; \; \; {\rm for}\; \; k > N \;.
\label{constraint}
\end{equation}
Given this constraint the dynamics has
$N$ true degrees of freedom.  The equation of motion for
$\dot \phi_k$ with $k > N$ becomes the following
equation of
constraint for the conjugate momenta:
\begin{equation}
\pi^*_k = i Q \sum_{j \le N} A_{k,j} \phi_j  \;. \label{picon}
\end{equation}
\vspace*{.12in}

Thus, even though $\phi_j$ is zero for $j>N$, $\pi_j$ is in
general non-zero; these non-zero conjugate momenta lead
directly to electric seagulls in the effective $N$ dimensional
Hamiltonian.
This is straightforward to see.  The simplest way is to
insert the constraint Eq.~(\ref{picon}) into the Hamiltonian.
The resulting collective Hamiltonian, $H^{\rm coll}$, reads
\begin{equation}
H^{\rm coll}  =  \sum_{i,j \le N}
\left [  \pi^*_i \delta_{i,j} \pi_j
\, +  \,  \phi^*_i B_{i,j} \phi_j  \, + \, i
Q (\pi^*_i A^*_{i,j} \phi^*_j - \pi_i A_{i,j} \phi_i) -
\sum_{k>N} Q^2 \phi^*_i A_{i,k} A^*_{k,j} \phi_j \right ] .
\\[.18in]
\end{equation}

\vspace*{.12in}
The last term is the electric seagull.
Of course, it is generally not permissible  to insert a solution
of the equation of motion back into the Hamiltonian since a
variation of the substituted Hamiltonian will not necessarily
give the same equation of motion as the original. However it is
trivial to show in this case that  the equations of motion for
the
collective Hamiltonian exactly reproduce
the equations of motion for the full Hamiltonian for the $N$
collective degrees
of freedom provided the noncollective degrees of freedom satisfy
the constraint Eqs.~(\ref{constraint}) and (\ref{picon}).

Thus we have shown explicitly how constraining the theory
leads to an effective electric seagull term.
Moreover, it is clear that the seagull term is physically
sensible in that it corresponds to an induced
electric charge density in the presence of a background field.
The charge density can be written as $\rho (x) = -i Q (\pi^*
\phi^* - \pi \phi)$.  This can be re-expressed as


\begin{eqnarray}
\rho(x) & = &\rho_1(x) + \rho_2(x)\\ [.18in]
\rho_1(x) & = & -i Q \sum_{i \le N,\:\: j \le N}
 f_i(x) f_j^*(x) \pi_i^* \phi_j^* + {\rm h.c.} \\[.18in]
\rho_2(x) & = & -i Q \sum_{j \le N, \:\: k > N}
 f_k(x) f_j^*(x) \pi_k^* \phi_j^* + {\rm h.c.} = i Q^2 \sum_{i,j
\le N, \:\: k > N}
 f_k(x) f_j^*(x) A^*_{k,i} \phi_i \phi_j^* \nn\\
\end{eqnarray}

\noindent where the last equality follows from the equation of
constraint
for $\pi_k$.
Clearly $\rho_2 (x)$ is the induced charge due to the electric
seagull.  Indeed the seagull term can be expressed as
\begin{equation}
H^{\rm seagull} \, = \,\frac{1}{2} \, \int \, dV \, A(x)
\rho_2(x) \, .
\end{equation}

Although the preceeding argument explicitly demonstrates that
electric seagulls can arise from a constraint onto a
collective subspace, it should be noted that the procedure used
here is {\it not} the one used in the soliton model calculations.
It is useful to show that, when applied to the present case, the
methods used there
yield the same electric seagulls
in the Hamiltonian as the construction above.
As already mentioned, in soliton models the constraint
on the fields is imposed at the level of the original Lagrangian
thereby obtaining
a collective Lagrangian. The collective Hamiltonian
is then derived from such a collective Lagrangian by means of
Legendre transformation acting on the collective variables.
It is not surprising that this gives
identical results to the collective Hamiltonian above.
Expressed in terms of our functions $f_i$ the collective
Lagrangian
reads
\begin{eqnarray}
L^{\rm coll} & = & L_1  + L_2 \;,\\ [.18in]
L_1 & = & \sum_{i , j \le N} \left [
\dot \phi^*_i  \delta_{i,j} \dot \phi_j \, - \,
\phi_i^* B_{i,j} \phi_j - i Q (\phi_j^* A_{j,i} \dot \phi_i
\, + \dot \phi_j^* A_{j,i} \phi_i ) \right ] \;, \\  [.18in]
L_2 & = &  Q^2 \left[
\sum_{i,j,l \le N} A_{i,l}A_{l,j} \phi^*_i \phi_j +
\sum_{i,j \le N,\:\: k>N} A_{i,k}A_{k,j} \phi^*_i \phi_j \right]
\end{eqnarray}

It is a simple exercise to show that the Legendre transform  of
$L^{\rm coll}$ yields $H^{\rm coll}$. It is worth observing
that the seagull term in $H^{\rm coll}$ is only (minus)
the second term in $L_2$. Thus, {\it in general} the procedure
used in
the soliton model calculations of Refs.
\cite{Che87}--\cite{GS93}---of
identifying the seagull in the collective Hamiltonian with minus
the seagull in the collective Lagrangian---is not correct.
It is only valid if the first term in $L_2$
vanishes, which generally
does not happen.  However, it is also worth
noting that symmetry can make such terms vanish for certain
classes of external electrostatic potentials.  For example, if
the $A_0$ distribution is odd under parity and all of the states
in the collective subspace have the same parity (either even or
odd) then $A_{i,j}=0$ for $i,j$ members of the collective
subspace and the first term in $L_2$ vanishes.   This is
significant  since this  is precisely  the circumstance of the
soliton models in a constant external electric field. The
electrostatic potential is just $- E_0 z$,
so $A_0$ is clearly odd; while all of the configurations in the
collective subspace in the Skyrmion, hedgehogs, and rotated
hedgehogs are odd under $x \rightarrow -x$ (making them even
under parity since under parity $x \rightarrow -x$ and $\pi
\rightarrow -\pi$) .

Thus, although  the procedure usually used in soliton models of
identifying the seagulls in the collective  Hamiltonian with the
negative of
the seagulls in the collective Lagrangian is not generally
valid, it is valid in the context in which it is actually used.
At this stage, it is worth noting that the objection of L'vov to
the procedure used in Refs.\cite{Che87}--\cite{GS93}
based on the case of a constant $A_0$ does not apply to the case
of constant electric field.
The procedure of Refs. \cite{Che87}--\cite{GS93}
should not be used for a constant $A_0$
since symmetry
does not
cause the second term of $L_2$ to vanish. In fact, it is rather
easy
to see that for constant $A_0$ it is the second term in $L_2$
that vanishes. This, of course, leads to a vanishing seagull
term in $H^{\rm coll}$ as is expected from gauge invariance.

Having shown how seagull terms can be induced by constraining
the allowed field configurations in the case of the simple
system of charge scalars we will study now how this mechanism
works for the actual case of soliton models.

\section{Electric seagulls in the Skyrme model}

As discussed in the Introduction from the point of view of
the electric seagulls the various versions of soliton model
behave in a very similar way. In this section we will
discuss in detail the particular case of the original Skyrme
model \cite{Sky61} with quartic term stabilization. However,
all our conclusions can be immediately extended to other
models based on hedgehog solitons.

We start by introducing the canonical pion field $\Phi_a$ in
terms of which the $SU(2)$ chiral field $U(x)$ reads
\be
U = {1\over{f_\pi}} \ \left( \Phi_0 + i \tau_a \Phi_a \right)
\ee
\nsp
where
\be
\Phi_0^2 = f_\pi^2 - \sum_{a=1}^3 \ \Phi_a^2 \;.
\ee
In terms of these fields the Hamiltonian of the Skyrme model
in the presence of an external e.m. field can be expressed
as \cite{SU94}
\be
{\tilde H_{Sky}} = H_{Sky} + H_1 + H_2 \;,
\label{hami}
\ee
where
\bea
H_{Sky} &=& \int dV \ \left\{ {1\over2} \ \Pi_a K_{ab}^{-1} \Pi_b
                                + \nu \right\} \;,\\ [.18in]
H_1 &=& \int dV \ eA_0 \left\{ -\rho -
                               J^0_a K_{ab}^{-1} \Pi_b \right\}
\;,
\\ [.18in]
H_2 &=& {1\over2} \int dV \ e^2 A_0^2 \left\{- \Gamma^{00} +
                            J_a^0 K^{-1}_{ab} J_b^{0} \right\}
\;.\\ [.18in] \nonumber
\eea
Here, $\rho$ is the electric charge density
and $\nu = \nu(\Phi_a,\nabla\Phi_a)$ are terms containing no time
derivative. The explicit form of these quantities can be found
{\it e.g.}, in Ref. \cite{Adk85}. $\Gamma^{00}$ is the seagull
term in
the
Lagrangian as defined in Ref. \cite{SU94}, and
\bea
\Pi_a &=& K_{ab} \dot \Phi_b + e A_0 J^0_a  \;,\\ [.18in]
J_a^0 &=& \epsilon_{3cd} \Phi_c K_{da} \;,\\ [.18in]
K_{ab}&=& X_{ab} -
      {1\over{e^2 f_\pi^4}} \left[ X_{ab} X_{cd}-X_{ac} X_{bd}
\right]
                   \partial_i \Phi_c \partial^i \Phi_d \;,\\
[.18in]
X_{ab}&=&\delta_{ab} + {\Phi_a \Phi_b\over{\Phi_0^2}} \;.
\eea

In writing Eq. (\ref{hami}) we have considered, as in the
previous
section,
the particular case in which the external magnetic field is zero
and have
chosen
a gauge in which $A_i=0$. Moreover, we have dropped several terms
which are not relevant for our arguments\footnote{They lead to
higher order corrections in $1/N_c$ once the rotating Skyrme
ansatz,
Eq. (\ref{ans}), is introduced \cite{SU94}.}.

As discussed in Ref. \cite{SU94}, explicit calculation shows that
the two terms in $H_2$ cancel each other. This is in complete
analogy
to what we have already seen in the previous section. At the
level of
the full Hamiltonian there are no electric seagull terms.
However,
within the Skyrme model, baryons are described in the large $N_c$
limit
as slowly rotating hedgehog configurations, namely, by field
configurations given by \cite{ANW83}
\bea
\Phi_a &=& f_\pi \ \sin F(r) \ R_{ai} \ \hat r_i  \;,\\ [.18in]
\Phi_0 &=& f_\pi \ \cos F(r) \;.
\label{ans}
\eea
$F(r)$ is the so-called chiral angle and
$R_{ab}$ is the time-dependent rotation matrix. The introduction
of this ansatz implies a strong restriction in the allowed
configurations
in the $\Phi$-field Hilbert space. As in the case of a charged
scalar field
this will lead to the appearance of an electric seagull term in
the
restricted (collective) Hamiltonian.

Replacing the Skyrme ansatz in $\tilde H_{Sky}$, Eq.
(\ref{hami}),
we get
\begin{equation}
\tilde H_{Sky}^{\rm coll} = M_{sol} + {1\over2} \Theta \Omega^2
-                 \int dV\ e A_0 \ \rho(r) -
                  {3\over{16\pi}} \int dV\ e^2 A_0^2 \ \Theta(r)
                  \left[ 1- R_{3n} R_{3p} \hat r_n \hat r_p
\right]\;.
\end{equation}
$M_{sol}$ is the soliton mass (see, {\it e.g.}, Ref. \cite{ANW83}
for its explicit expression), $\vec \Omega$ is the angular velocity
defined
by $R_{an} \dot R_{ai} =
\epsilon_{nil} \Omega_l$, the charge density $\rho(r)$ is given
by
\be
\rho(r) = -{1\over{4\pi^2}} {\sin^2 F\over{r^2}} F' \;,
\ee
and $\Theta$ is the soliton moment of inertia
\be
\Theta = \int dr \ r^2 \ \Theta(r)
\ee
where
\be
\Theta(r) = {8\pi f_\pi^2\over3} \ \sin^2F
             \left[ 1 + {1\over{e^2 f_\pi^2}}
                    \left( F'^2 + {\sin^2 F\over{r^2}} \right)
\right] \;.
\ee
\vspace*{.15in}
\noindent As we see, although $H_2$ vanished, we do have terms
proportional
to $e^2$---{\it i.e.}, seagull-like type---in $\tilde
H_{Sky}^{\rm coll}$.
They have their origin in $H_{Sky}$ and $H_1$.
In order to find the canonical expression of the collective
Hamiltonian we have still to eliminate the angular velocity
$\Omega_i$
in
terms of the angular momentum operator $J_i$. The general form of
this
operator can be obtained from the full Skyrme model Lagrangian in
the
presence of an external $A_0$ field by using
\be
J_i = \int dV \ \epsilon_{ijk}\ r_j \ T_{0k} \;,
\ee
where $T_{0k} = \Pi_a \partial_k \Phi_a$ are the corresponding
components
of the energy--momentum tensor. We get
\be
J_i = \int dV \ \epsilon_{ijk} \ r_j \ K_{ab}\
        \dot \Phi_b \partial_k \Phi_a
      + \int dV \ eA_0 \ \epsilon_{ijk}\ \epsilon_{3cd}\ r_j\
       K_{ab}\ \Phi_c \partial_k \Phi_a \;.
\ee
The collective form of this operator is then obtained by
replacing the Skyrme ansatz, Eq. (\ref{ans}). We
obtain
\be
J_i = \Theta \Omega_i + {3\over{8\pi}}
       \int dV \ eA_0 \Theta(r)
        \left( \delta_{in} - \hat r_i \hat r_n \right) \ R_{3n}
\;.
\label{jota}
\ee
Using this equation, we get the final expression of the
Hamiltonian
in the restricted space of the rotating soliton
\be
\tilde H_{Sky}^{\rm coll} = H_0 + H_L + H_Q \;,
\ee
where
\bea
H_0 &=& M_{sol} + {J^2\over{2\Omega}} \;,\\ [.18in]
H_L &=& - \int dV eA_0 \left[ \rho(r) + {3\over{8\pi\Theta}}
        \Theta (r)
        \left( \delta_{in} - \hat r_n \hat r_i \right) \ R_{3n}
J_i \right] \;, \\ [.18in]
H_Q \! &=& \! {\Theta\over2} \left( {3\over{8\pi\Theta}} \right)^2
        \left[ \int dV e A_0 \Theta(r)
        \left(\delta_{in} - \hat r_i \hat r_n \right) \right]
        \left[ \int dV e A_0 \Theta(r)
        \left(\delta_{im} - \hat r_i \hat r_m \right) \right]
         R_{3n} R_{3m}  \nonumber \\ [.18in]
    & & \qquad \qquad - {3\over{16\pi}} \int dV e^2 A_0^2
\Theta(r)
         \left[ 1 - R_{3n} R_{3m} \hat r_n \hat r_m \right] \;.
\eea

We can now discuss some particular cases. First, we will
consider the situation proposed by L'vov in which $A_0$ is a
constant
field. In this case it is easy to show that
\bea
 {3\over{8\pi\Theta}}
         \int dV e A_0 \Theta(r)
        \left( \delta_{in} - \hat r_i \hat r_n \right)
   &=& e A_0 \delta_{ni} \;,\\ [.18in]
{3\over{16\pi}} \int dV e^2 A_0^2 \Theta(r)
         \left[ 1 - R_{3n} R_{3m} \hat r_n \hat r_m \right]
   &=& {\Theta\over2} e^2 A_0^2 \;.
\eea
from where we get
\bea
H_L(A_0=cte) &=& - e A_0 \left({1\over2} + I_3 \right) \;,\\
H_Q(A_0=cte) &=& 0 \;.
\eea
Here we have used the well-known relation $I_3 = R_{3i} J_i$ and
$\int dV \rho(r) = 1/2$. As expected in the case of a constant
$A_0$ field
the quadratic term in $A_0$ vanishes and the resulting
Hamiltonian
corresponds to the interaction of the constant $A_0$ field with
the baryon
electric charge.

Now we turn to the more interesting case of a constant electric
field
in the $z$-direction, namely $A_0 = - z E$. This is the
field configuration that has been
used in Refs. \cite{Che87}--\cite{GS93}
to determine the electric polarizabilities. It is easy to show
that
in this case
\bea
 \int dV \ eA_0 \ \rho(r) &=& 0 \;,\\ [.18in]
 {3\over{8\pi\Theta}}
         \int dV e A_0 \Theta(r)
        \left(\delta_{in} - \hat r_i \hat r_n \right)
   &=& 0 \;,\\ [.18in]
{3\over{16\pi}} \int dV e^2 A_0^2 \Theta(r)
         \left[ 1 - R_{3n} R_{3m} \hat r_n \hat r_m \right]
   &=& {e^2 E^2\over6} \int dr r^4 \Theta(r)
         \left[ 1 - {2\over5} D_{00}^{(2)} \right] \;,
\eea
where we have used $(R_{33})^2 = {1\over3} + {2\over3}
D_{00}^{(2)}$
with $D_{00}^{(2)}$ being the corresponding $D$-matrix. Using
these
relations we get
\bea
H_L(E=cte) &=& 0 \;,\\
H_Q(E=cte) &=& - {1\over2} \ \gamma_e \ E^2
               \left[ 1 - {2\over5} D_{00}^{(2)} \right] \;,
\eea
where
\bea
\gamma_e &=& {e^2\over3} \int dr \ r^4 \ \Theta(r)\;.
\eea
These expressions coincide with the ones usually used
to calculate the electric polarizabilities in the Skyrme model
(see {\it e.g}, Ref. \cite{SM92}). In fact, for the case of the
nucleon
the matrix element of $D_{00}^{(2)}$ vanishes and we simply get
\be
\alpha_N = \gamma_e \;.
\ee

As already found in the case of the charged scalars, although it
is not
in general correct to take the interaction terms in the
Hamiltonian as
minus those in the Lagrangian, this relation does hold for the
case of a
constant electric field.

\section{Conclusions}

In conclusion we have shown that although a fundamental bosonic
theory has no electric seagulls at the Hamiltonian level the act
of constraining the  dynamics to a collective subspace induces
electric seagulls in the collective Hamiltonian. We have
explicitly
shown
how this mechanism works in a simple model of a charged scalar
and
in the more relevant case of the Skyrme model. We have
argued that this can be immediately generalized to other versions
of
soliton models. Moreover, although we have only discussed how
these seagull terms contribute to the nucleon electric
polarizabilities,
it is clear that our conclusions can be extended to other baryons
as the
$\Delta$ and strange hyperons. We have demonstrated that the
correct
collective Hamiltonian can be obtained either by imposing the
constraint
on the Lagrangian to obtain a collective Lagrangian and then
taking a
Legendre transform, or by first obtaining the full Hamiltonian
and
then
applying the field constraint; both methods leading to the same
result.
We have also discussed that the Legrendre transform for the
seagull term in the Lagrangian is simply given by the negative of
the term
provided certain symmetry conditions are met. In particular, for
the case
of a constant electric field such conditions are fulfilled. Thus,
we
conclude that the procedures
used in  Refs. \cite{Che87}--\cite{GS93}
to
determine the seagull contributions to the electric
polarizabilities are
completely valid. On the other hand, for constant $A_0$ this
relation does not hold. Explicit calculation shows, as expected,
that
electric seagulls in $H^{\rm coll}$ vanish in that case.

We should also mention the interesting work
of Nikolov, Broniowski and Goeke \cite{NBG94}.  This work is
based on a Nambu--Jona-Lasinio type model. As the model has no
fundamental pion, it manifestly has no fundamental $\pi \gamma
\gamma$ seagull terms. Therefore, their calculation is done in
terms
of the fundamental quark degrees of freedom which, of course,
give only
dispersive contributions. However,
they also show that if one wishes to obtain a Skyrme-type model
with pions from the NJL model via a gradient expansion, the
leading term in the expansion for the electric polarizability
is precisely the seagull term of the $\sigma$ model approaches.
Again we see the general point: an effective model written in
terms
of meson fields restricted to a collective manifold does
contain electric seagulls.

Finally, we want to mention that in general the seagull terms
considered here are not the only contributions to the electric
polarizabilities. In fact, when pion fluctuations around the
soliton
are taken into account the linear terms $H_L$ can contribute in
second order perturbation theory. For the case of the nucleon the
corresponding intermediate states are dipole-excited,
negative-parity, nucleon resonances. A rough estimation
\cite{Che87} shows that these contributions are small.
Similar result is found for the case of the
$\Lambda$ \cite{GSS95}.
Thus, the total electric polarizability is dominated by the
seagull term--as
expected by naive $N_c$ counting--and the arguments based in
chiral
perturbation theory given in the Introduction. Another point to
be mentioned
is that the techniques discussed in this article can also be
applied for the
case of a constant magnetic field, {\it i.e.}, in the
determination of the diamagnetic contributions to the
magnetic polarizabilities. In that case, as already found in
Ref. \cite{SW89},
the seagull terms in the Lagrangian differ from those in the
Hamiltonian.
Numerically, however, this difference is negligible.

\section*{Acknowledgements}

We wish to thank N. Kaiser, B. Schwesinger and W. Weise for
useful
discussions. Most of the work reported here was done at the
program on ``Chiral Dynamics in Hadrons
and Nuclei" (INT-95-1) at the Institute for Nuclear Theory at the
University of
Washington. We  thank the organizers of the program for
inviting us and the U.S. Department of Energy
for partial financial support during that period.  One of us
(TDC)  acknowledges additional financial support of the U.S.
Department of Energy under grant no. DE-FG02-93ER-40762 and the
U. S. National Science Foundation under grant no. PHY-9058487.
He also  thanks the Institute for Nuclear Theory and the
Department of Physics at the University of Washington for their
hospitality during an extended stay.

\vfill
\pagebreak


\vfill
\pagebreak

\end{document}